\documentclass[superscriptaddress,showpacs,amsmath,amssymb,reprint,nobibnotes,nofootinbib,aps,prb]{revtex4-1}

\usepackage{graphicx}
\usepackage{bm}
\usepackage[colorlinks=true,linkcolor=blue,citecolor=blue,urlcolor=blue]{hyperref}

\begin{document}

\title{Limiting mechanism for critical current in topologically frustrated Josephson junctions}
\author{Sarah B. Etter}
\email{etters@itp.phys.ethz.ch}
\affiliation{Institute for Theoretical Physics, ETH Zurich, 8093 Zurich, Switzerland}
\author{Hirono Kaneyasu}
\affiliation{Department of Material Science, University of Hyogo, Kamigori, Hyogo
678-1297, Japan}
\author{Matthias Ossadnik}
\author{Manfred Sigrist}
\affiliation{Institute for Theoretical Physics, ETH Zurich, 8093 Zurich, Switzerland}
\date{\today}

\begin{abstract}
Eutectic Sr$_2$RuO$_4$-Ru samples with $\mu$m-sized Ru-metal inclusions support inhomogeneous superconductivity above the bulk transition of Sr$_2$RuO$_4$ in the so-called 3-Kelvin phase. In Pb/Ru/Sr$_2$RuO$_4$ Josephson junctions as realized by Maeno \textit{et al.}, a Pb film is indirectly coupled to the superconductor Sr$_2$RuO$_4$ mediated by the proximity-induced superconducting Ru-inclusions, yielding an extended Josephson contact through the interface between Ru and Sr$_2$RuO$_4$. Motivated by this experimental setup, we formulate a sine-Gordon model for the Josephson phase of the interface, assuming a simple cylindrical shape for the Ru-inclusion hosting the proximity-induced $s$-wave superconducting phase. Considering the Sr$_2$RuO$_4$ as a chiral $p$-wave superconductor, we discuss two types of Josephson junctions, a frustrated one due to the nature of the order parameter in Sr$_2$RuO$_4$, and an unfrustrated one for the topologically trivial 3-Kelvin phase. While the latter situation displays standard junction behavior, the former  yields an unusual limiting mechanism for the critical current, based on a pinning-depinning transition of a spontaneously induced magnetic flux driven by an externally applied current. We analyze different coupling limits and show that different critical currents can arise for the two topologies. This concept fits well to recent experimental data obtained for the above setup showing an anomalous temperature dependence of the critical current at the transition temperature $T_c $ of bulk Sr$_2$RuO$_4$.  
\end{abstract}


\maketitle


\section{Introduction}

Since its discovery nearly two decades ago the superconducting transition metal oxide Sr$_2$RuO$_4$ has been one of the most extensively investigated low-temperature superconductors \cite{maeno:1994,mackenzie:2003,maeno:2012}. Numerous studies provide strong evidence for unconventional pairing,
with the chiral $p$-wave state emerging as one of the most promising candidates, a topological superconducting phase \cite{maeno:2001}. The results of zero-field $\mu$SR experiment indicating a time-reversal symmetry breaking superconducting phase are consistent with this phase \cite{luke:1998} as well as the observation of 
a polar Kerr effect \cite{xia:2006}. The existence of chiral domains is suggested by Josephson interference experiments \cite{kidwingira:2006}. 
One feature characteristic for the chiral $p$-wave state is the presence of topologically protected chiral edge states which carry a spontaneous supercurrent. Experiments directly aiming at the detection of magnetic fields due to the chiral edge currents yield a  negative result, casting some doubt on the identification of the pairing symmetry \cite{kirtley:2007,kallin:2012}. On the other hand, the presence of edge states has been firmly reported by several groups \cite{liu:2003,laube:2000,kashiwaya:2011}. 

In addition, topology plays a role in the overall phase coherence of the superconducting state, as has been suggested recently in the context of eutectic Sr$_2$RuO$_4$-Ru \cite{kaneyasu:2010a}. In such samples excess Ru segregates from the bulk material to form $\mu$m-size Ru-metal inclusions in the Sr$_2$RuO$_4$ matrix. In this environment superconductivity appears as an inhomogeneous phase, the so-called \mbox{`3-Kelvin'} phase, already at an onset temperature of $T^* \approx 3~$K which is twice as high as the bulk superconducting transition temperature $T_c \approx 1.5~$K \cite{maeno:1998}. It has been proposed that superconductivity nucleates at the interfaces between Ru and Sr$_2$RuO$_4$ \cite{sigrist:2001,yaguchi:2003}. Assuming that the bulk superconducting phase has chiral $p$-wave symmetry, one finds that the structure of the {3-K} phase is different and thus an additional transition beyond mere percolation has to occur on the way to full coherence throughout the system. This additional transition involves both time reversal symmetry breaking and the change of overall topology. 
Several experiments give hints of such an additional transition within the {3-K} phase between $T^*$ and $T_c$ \cite{mao:2001,hooper:2004,kawamura:2005}. 

\begin{figure}[htb]
\includegraphics[width=0.8\columnwidth]{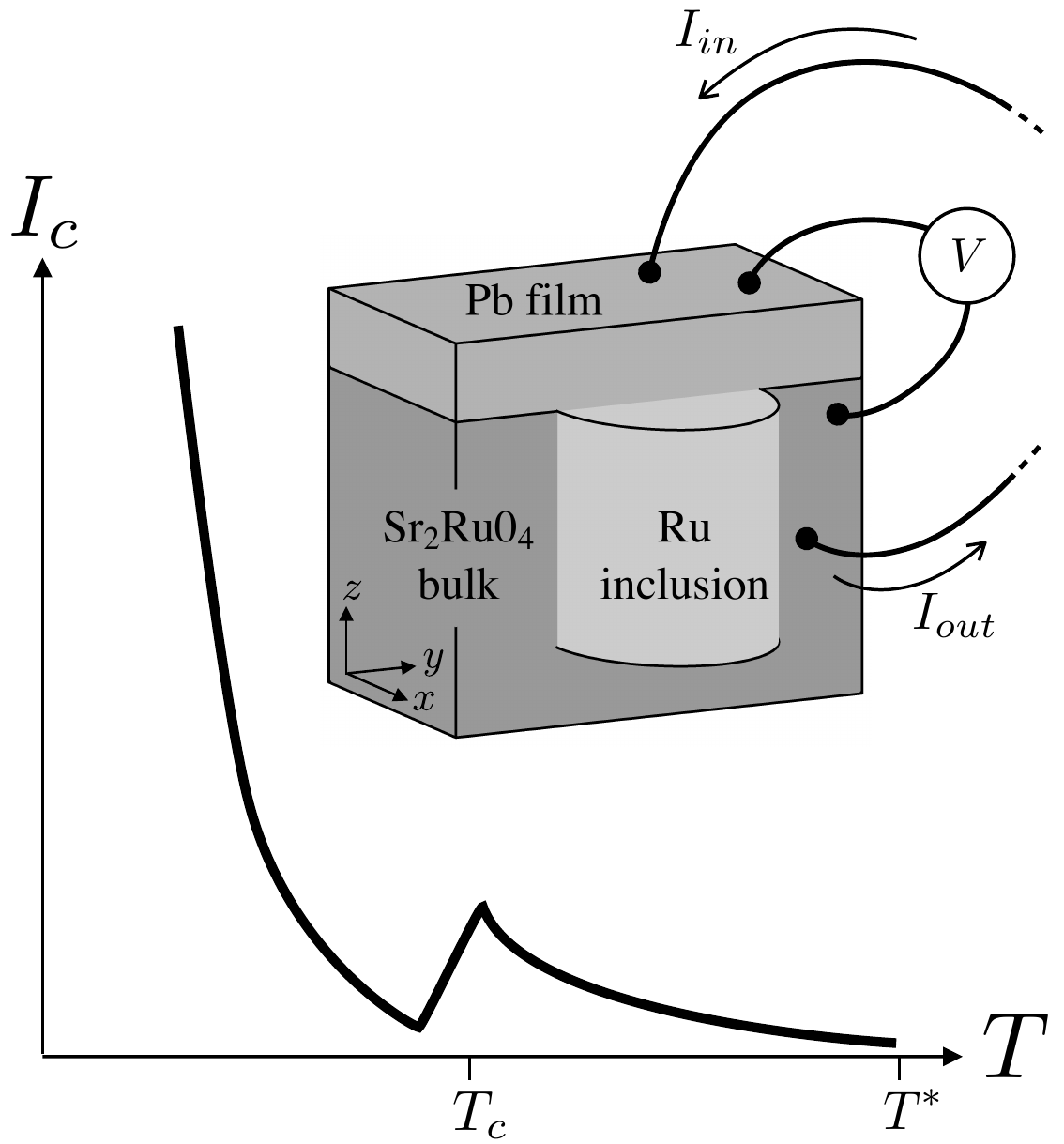}
\caption{\label{fig:schematic} Sketch of the anomalous behavior of the critical current based on the results by Maeno \textit{et al.} \cite{nakamura:2011} and a schematic of the Pb/Ru/Sr$_2$RuO$_4$ setup. Below the onset of the 3-Kelvin phase at $T^*$, the critical current first increases as expected upon lowering the temperature, but suddenly drops around the bulk critical temperature of Sr$_2$RuO$_4$, $T_c$, after which it recovers and increases rapidly. This effect has been studied in a Pb/Ru/Sr$_2$RuO$_4$ Josephson junction where an external current is applied to the eutectic Sr$_2$RuO$_4$-Ru through a Pb film and the voltage is measured across the junction.}
\end{figure}

Another topological feature occurs in connection with the Josephson effect. Assume that the Ru-inclusion itself is superconducting with conventional $s$-wave symmetry. What are the consequences? In recent experiments, Maeno \textit{et al.} \cite{nakamura:2011} have realized such a situation by depositing Pb on top of a eutectic Sr$_2$RuO$_4$-Ru sample, as shown in the inset of Fig.~\ref{fig:schematic}. In this way they created a device which functions as a Pb/Ru/Sr$_2$RuO$_4$ Josephson junction since, by proximity, the superconductivity of Pb penetrates the Ru-inclusion while there is little penetration of this superconductivity directly into Sr$_2$RuO$_4$ due to its electronic structure, i.e.\ the very weak dispersion along the c-axis \cite{mackenzie:2003}. Moreover, the Josephson coupling between an $s$-wave state and the chiral $p$-wave state along the c-axis is rather weak for symmetry reasons\cite{geshkenbein:1986}. In addition to the Josephson coupling between Pb and Sr$_2$RuO$_4$, which does not exist through direct contact alone, but is mediated by the Ru-inclusion, they also observed an anomalous temperature dependence of the Josephson critical current, as illustrated in Fig.~\ref{fig:schematic}. First, a Josephson current already appears at a temperature above $T_c$ and below $T^*$, indicating that most likely coupling to the {3-K} phase is observed here. The critical current $I_c$ increases with lowering temperature in this regime. Second, $I_c$ rather suddenly drops around $T_c$, but then recovers quickly and increases further at lower temperatures.

While the anomalous behavior of the critical current in this device is an interesting topic that we will discuss in detail elsewhere, we would like to show that in the temperature regime below $T_c$ there is an unusual limiting mechanism for the Josephson critical current at the Ru-Sr$_2$RuO$_4$ interface. This originates from a pinning-depinning transition of a spontaneous flux pattern on the interface. The Josephson coupling between the inclusion supporting $s$-wave superconductivity and the surrounding chiral $p$-wave bulk relies on the effect of spin-orbit coupling in order to provide lowest order pair tunneling. It has been shown recently that the structure of this coupling leads to a Josephson phase between the two superconductors which is frustrated\cite{kaneyasu:2010}. A magnetic flux pattern emerges in order to release this frustration. On the other hand, the state that nucleates at $T^*$ has a different structure and does not lead to frustration, but is optimally coupled to the $s$-wave inclusion and is therefore subject to the standard limiting of Josephson current. 

In this article we will consider a single, cylindrical Ru-inclusion inside the Sr$_2$RuO$_4$ bulk since this illustrates the mechanism limiting the Josephson critical current most clearly. We introduce our model in Sec.~\ref{sec2}, formulating the Josephson effect through a sine-Gordon equation which allows us to discuss both the frustrated and unfrustrated situation within the same framework. In Sec.~\ref{sec3} we present the different junction behaviors and compare them. We start by reviewing the situations discussed before. We then demonstrate that in the frustrated case a completely rotationally symmetric inclusion does not support a supercurrent because the spontaneous magnetic flux is driven through the applied current, generating a voltage. An inhomogeneous junction on the other hand, implemented by allowing the coupling strength to vary along the interface, leads to pinning effects giving rise to a finite Josephson current. In Sec.~\ref{sec:pin-strong} we continue with a detailed discussion of this pinning-depinning mechanism and how it limits the critical current.


\section{Model}\label{sec2}

\begin{figure}[htb]
\includegraphics[width=0.7\columnwidth]{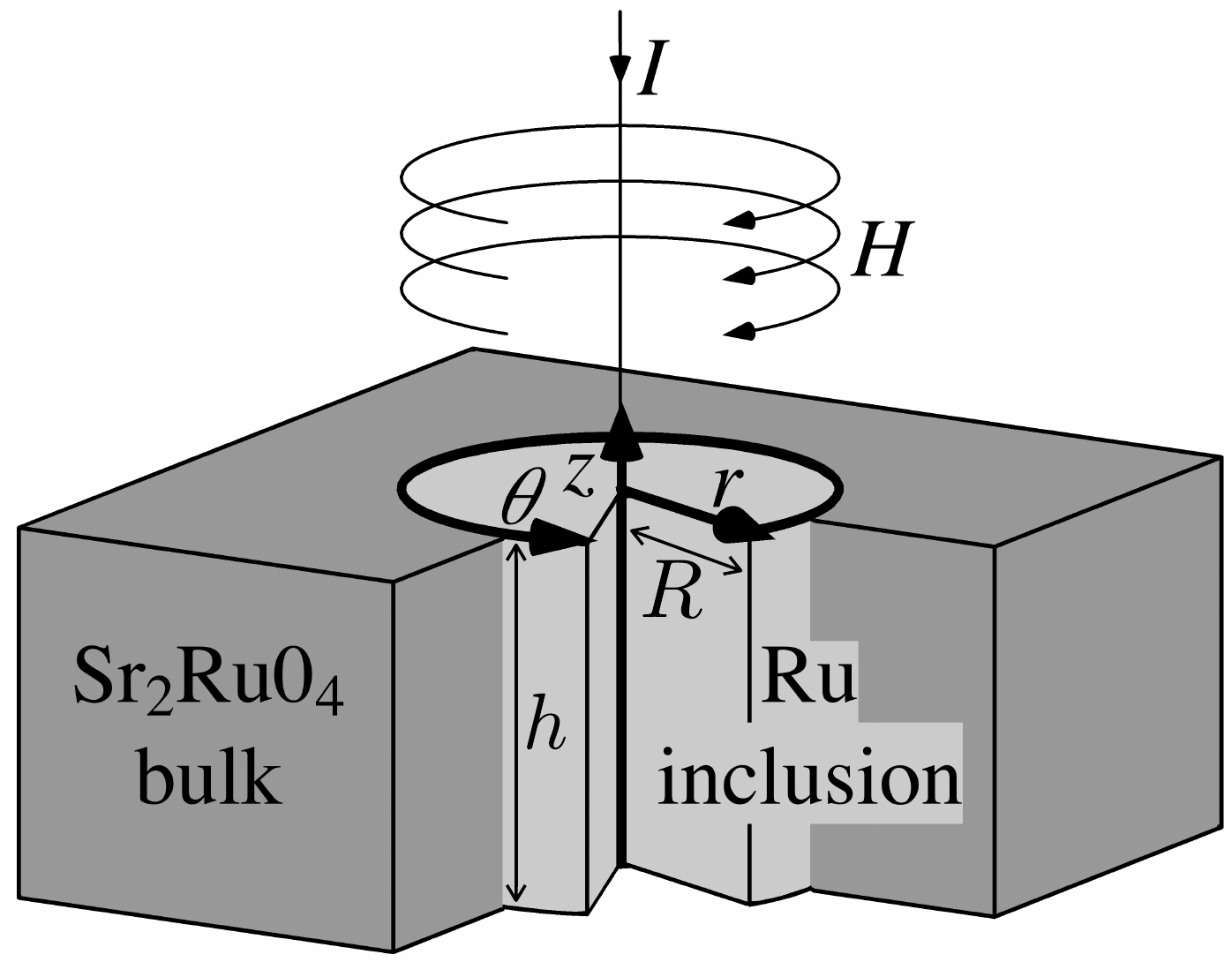}
\caption{\label{fig:model} Our model of the Pb/Ru/Sr$_2$RuO$_4$ Josephson junction. We consider a single Ru inclusion of cylindrical shape with radius $R$ and height $h$, surrounded by the Sr$_2$RuO$_4$ bulk. The contact through the Pb film is approximated by an external current $I$ applied along the $z$-axis at the top which is equivalent to an external circular magnetic field $H$.}
\end{figure}

We obtain a simple and illustrative model of the Pb/Ru/Sr$_2$RuO$_4$ Josephson junction by considering a single Ru-inclusion in the shape of a cylinder of height $h$ and radius $R$, surrounded by the Sr$_2$RuO$_4$ bulk, as illustrated in Fig.~\ref{fig:model}. The axis of the cylinder is parallel to the $c$-axis of the tetragonal crystal of Sr$_2$RuO$_4$. This is also the axis of chirality for the chiral $p$-wave state denoted as
\begin{equation}
\bm{d} (\bm{k}) =  \Delta_p \hat{\bm{z}} (\hat{k}_x \pm i \hat{k}_y ) = \hat{\bm{z}} \bm{\eta} \cdot \hat{\bm{k}},
\end{equation}
where $\Delta_p$ is the complex gap amplitude that generally depends on space and temperature. The orientation of the $\bm{d}$-vector along the $z$-direction means equal-spin-pairing in the basal plane of the tetragonal lattice. Note that this state is two-fold degenerate and has two order parameter components $\bm{\eta} = \eta_x \hat{\bm{x}}+ \eta_y\hat{\bm{y}}$ which in the bulk phase are given by $\bm{\eta} = \Delta_p (1, \pm i)$. 

We now consider the interface between the cylindrical $s$-wave superconductor with order parameter $\eta_s$ and the bulk $p$-wave superconductor as an extended Josephson junction. Focusing on the Josephson effect we assume that the order parameters as such are rigid, but let the Josephson phase between the two superconductors vary along the interface. In the following we develop an effective model for this situation based on the free energy of the junction
\begin{equation}
\mathcal F_{int}=d\int_0^{2\pi}\!R\text{d}\theta\int_{0}^h\!\text{d}z\,f_{int}(\theta,z),
\end{equation}
with $d$ the effective width, $d=\lambda_{Ru}+\lambda_{bulk}+d_0$, where $\lambda_{Ru,bulk}$ are the London penetration depths of the two materials and $d_0$ is the actual width of the interface. The free energy density takes into account that the even-parity spin-singlet $ \eta_s $ and odd-parity spin-triplet order parameter $ \bm{\eta} $ have to be matched through spin-orbit coupling. Thus, by symmetry, it has the following form
\begin{equation}
f_{int}(\theta,z)= \frac{\bm{B^2}}{8\pi}-\frac{K}{d}\left[ \eta_s^*\left(\bm{\hat z}\cdot(\bm{\hat n}\times\bm{\eta})\right)+\textrm{c.c.}\right],
\label{josephson-1}
\end{equation}
with $K$ a coupling constant and $\hat{\bm{n}} = \hat{\bm{r}}$ the normal vector of the interface in cylindrical coordinates \cite{geshkenbein:1986}. The first term contains the energy of the magnetic self-field $\bm{B}$ confined within the range $d$ and the second term is the Josephson coupling energy. 

To discuss the Josephson effect we first introduce the two phases of interest in Sr$_2$RuO$_4$. The phase nucleating at $T^*$ corresponds to the single component of the $ p $-wave state parallel to the interface, as was shown previously in Ref.~\onlinecite{sigrist:2001}. We call this the phase~A (see Ref.~\onlinecite{kaneyasu:2010a}) represented by the order parameter
\begin{equation}
\bm{\eta}^{(\text{A})} = \Delta_p \hat{\bm{z}} \times \hat{\bm{n}}.
\end{equation}
At a lower temperature the component perpendicular to the interface also nucleates but it does not couple to the Ru-order parameter to lowest order, as follows from Eq.(\ref{josephson-1}), and we neglect it. We also ignore the spatial dependence of the gap amplitude $\Delta_p$ perpendicular to the interface in this phase, using its value at the interface only. The order parameter of the bulk phase of Sr$_2$RuO$_4$, the phase~B, is in cylindrical coordinates given by
\begin{equation}
\bm{\eta}^{(\text{B})}(\theta) = \eta_r \hat{\bm{r}}+ \eta_\theta \hat{\bm{\theta}}= \Delta_p( 1,\pm i ) e^{iN\theta},
\end{equation}
where $N=\pm1$ is the winding number of the order parameter around the inclusion, relating to the two chiralities. Finally, fixing the global phase of the $p$-wave order parameter by choosing a proper gauge, we define the Josephson phase $ \phi(\theta,z) $  through the order parameter on the $s$-wave side,
\begin{equation}
\eta_s(R,\theta,z)  = | \eta_s | e^{ i \phi(\theta,z)},
\end{equation}
which is variable along the interface, while we keep $| \eta_s|$ constant as discussed above.

Using the standard procedure to derive the sine-Gordon model for the phase of an extended Josephson junction, we formulate for the free energy density
\begin{align}
 &f_{sg}(\theta,z)=\frac{2(2\pi)^3d^2}{\Phi_0^2}f_{int}(\theta,z) \label{eq:sgfree} \\ 
 &=\frac{1}{2} \left\{\left(\frac{\partial\phi}{\partial z}\right)^2+\left(\frac{\partial\phi}{R\partial\theta}\right)^2 \right\} -\frac{1}{\lambda_J^2} \cos [\phi(\theta,z)- N \theta ]\nonumber
\end{align}
with the Josephson penetration depth $\lambda_J$ defined as
\begin{equation}
\lambda_J^2=\frac{\Phi_0^2}{4(2\pi)^3dK|\eta_s||\bm{\eta}|}
\end{equation}
and with the winding number extended to both phases
\begin{equation}
N = \left\{ \begin{array}{ll}  0 & \quad\text{phase A}, \\ & \\
\pm1  & \quad\text{phase B}.
\end{array} \right.
\end{equation}
The local Josephson critical current density per area is given by
\begin{equation}
J_c = \frac{4 \pi c}{\Phi_0} K |\eta_s||\bm{\eta}|=\frac{c\Phi_0}{8\pi^2d}\frac{1}{\lambda_J^2}.
\label{Jc-lambda}
\end{equation}
By variation with respect to the phase $\phi(\theta,z)$ we obtain the differential equation
\begin{equation}
\frac{\partial^2\phi(\theta,z)}{R^2\partial\theta^2}+\frac{\partial^2\phi(\theta,z)}{\partial z^2}=\frac{1}{\lambda_J^2}\sin[\phi(\theta,z)-N\theta].
\label{eq:sg}
\end{equation}
The boundary conditions are determined by the externally applied current $I$, injected through the top of the Ru-cylinder ($I>0$). The cross-sectional current density is given by
\begin{equation}
\bm{J} = - \frac{I}{\pi R^2}  \hat{\bm{z}} \Theta(z-h) \Theta(R-r),
\end{equation}
The current can also be translated into a circular magnetic field (see Fig.~\ref{fig:model}) on top of the cylinder 
\begin{equation}
\bm{H} = -\frac{2I}{cR}\hat{\bm{\theta}} .
\label{eq:circmag}
\end{equation}
The connection between magnetic field and phase on the interface is given by the standard relation \cite{tinkham:2004}
\begin{equation}
\bm{B}(\theta,z) = \frac{\Phi_0}{2 \pi d} \left\{ \hat{\bm{n}} \times \bm{\nabla} \phi(\theta,z) \right\},
\end{equation}
which defines the local magnetic flux density confined in the range $d$ at the interface. Combining this relation with Eq.~(\ref{eq:circmag}) we write the boundary conditions for the upper ($z=h$) and lower ($z=0$) boundary,
\begin{subequations}\label{bc-phi}
\begin{align}
 \left. \frac{\partial \phi}{\partial z} \right|_{z=h} &= \frac{4 \pi d}{c\Phi_0 R} I = \gamma ,  \\ 
 \left. \frac{\partial \phi}{\partial z} \right|_{z=0} &= 0,
\end{align}
\end{subequations}
where, for future convenience, we abbreviate the expression of the first condition by the parameter $ \gamma $ in the following. The second condition corresponds to a vanishing current through the bottom of the cylinder, since we assume that all current leaves the Ru-inclusion through the interface (see below). The solution which we discuss in the following could be extended by mirror symmetry at the plane $ z=0 $ to a cylinder of height $ 2h $ with a current of the same magnitude entering the cylinder also through the bottom at $ z = -h $. Eventually, of course, we also have periodic boundary conditions along $ \hat{\bm{\theta}} $, i.e.\ $ \phi(\theta + 2 \pi,z) = \phi(\theta,z) $.

A comment is in order at this point. We assume that the current enters through the top of the cylinder only, as is anticipated from the experimental setup. We differ from the experiment by keeping the $s$-wave order parameter in the cylinder independent of $z$, while in reality the proximity effect from Pb through the top would yield a rather strongly $z$-dependent part. Also, Pb itself does not appear explicitly in our setup. However, these features of the model are not important for the discussion of the limiting of the Josephson effect through the interface which we consider the weakest link for supercurrent flow in the whole device. 


\section{Results: Different junction behaviors}\label{sec3}

We will first discuss the situation of a system with complete rotational symmetry around the cylinder axis, i.e.\ a Josephson coupling which is constant over the whole interface. This way we can easily explain the difference between the phases~A and B: the junction is unfrustrated or frustrated, respectively, depending on the phase winding $N$. We then change the setup to allow an inhomogeneous Josephson coupling varying along the $\theta$-direction. We show that this is essential for stabilizing a finite supercurrent in phase~B. 

\subsection{Phase A: Unfrustrated case}\label{sec:unfrust}

In phase~A, there is no phase winding and $N=0$ in the Josephson coupling part of Eqs.~(\ref{eq:sgfree}) and (\ref{eq:sg}). The junction is unfrustrated. Here, the standard discussion of an extended junction applies as described e.g.\ by Barone and Patern\`o \cite{barone:1982}. 
Although the nucleation of superconductivity in the 3-Kelvin phase is inhomogeneous, i.e.\ restricted to a layer at the interface of the thickness of roughly the coherence length, we ignore this aspect here and assume that both superconducting phases have bulk character, for simplicity. 
The existence of a Josephson coupling as detected in the experiment relies on the observation of the usual anomaly in the current-voltage relation of the junction, and indicates the existence of a superconducting condensate on both sides of the interface \cite{nakamura:2011}. Restricting our analysis to the interface only is sufficient when discussing the essential features of phase~A.

With the vanishing phase winding $N=0$, the phase depends only on $z$, i.e.\ $\phi(\theta,z) = \phi(z)$, and it is constant along $\theta$ with the periodic boundary conditions fulfilled automatically. The current flows radially through the interface, $\bm J=J_r(z)\hat{\bm r}$, yielding a circular magnetic field $ \bm{B} = B_{\theta}(z) \hat{\bm{\theta}} $ which equals $ \bm H $ at $ z =h $ and penetrates on the length scale of the Josephson penetration depth $ \lambda_J $. Our cylindrical geometry can be translated into a one-dimensional extended junction with two boundaries at $z=0$ and $h$. For this situation Owen and Scalapino have discussed the behavior of the junction\cite{owen:1967}. 

\begin{figure}[htb]
\includegraphics[width=0.95\columnwidth]{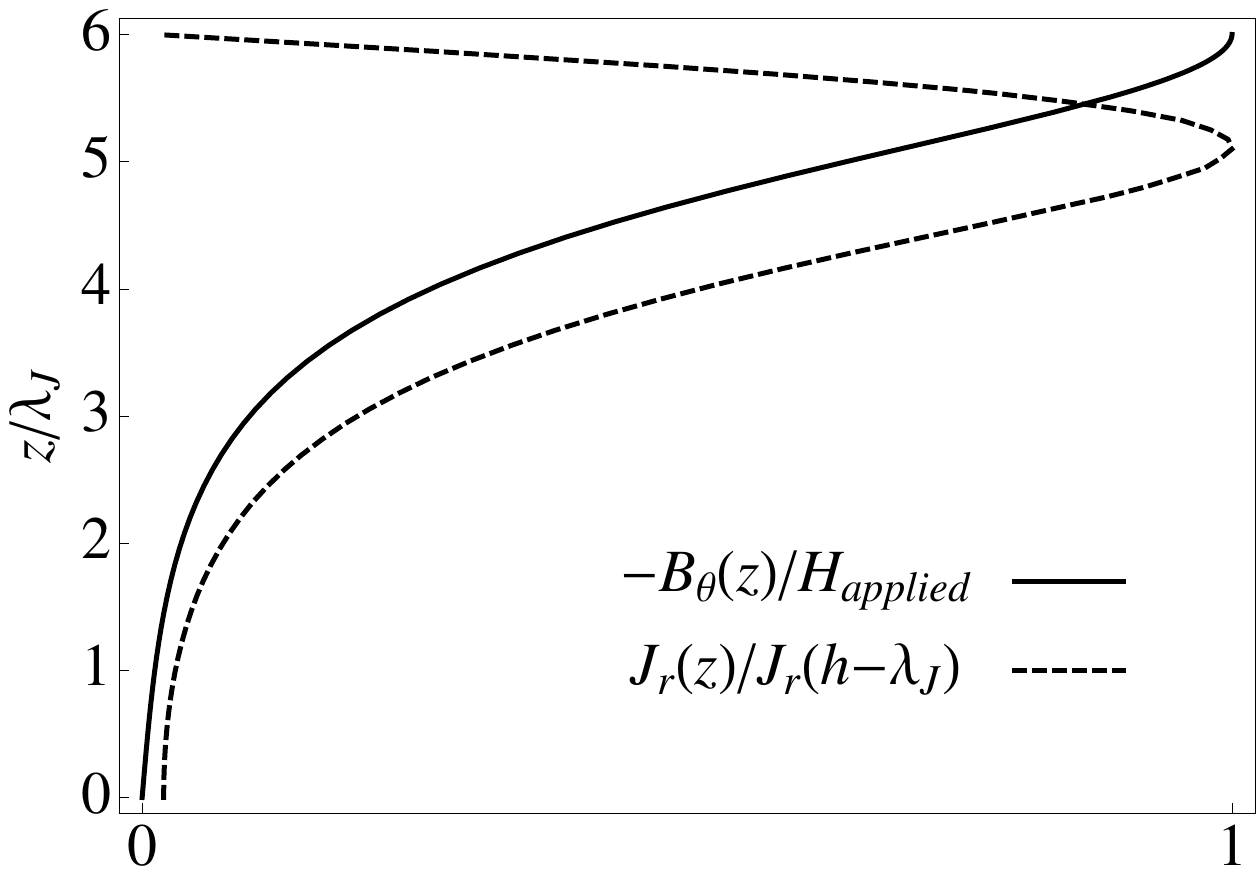}
\caption{\label{fig:unfrust} Flux pattern and radial current density in the unfrustrated case $N=0$ at the critical current $I=I_{c}$. The Josephson penetration depth is $\lambda_J=0.5 R$ and the height of the cylinder is $h=3 R$. The height in terms of the Josephson penetration depth is $h=6\lambda_J$. Both curves are normalized with respect to their maximum. The magnetic flux is maximal at the top edge where the external current is injected and it extends into the junction on the order of $\lambda_J$, while the current density peaks at this depth. These are the standard results for an extended Josephson junction.}
\end{figure}

Considering the case of a long cylinder, $ h \gg \lambda_J $, the current is concentrated at the top of the cylinder with a peak positioned roughly at the depth $ \lambda_J $, leading to a ring-shaped current pattern as shown in Fig.~\ref{fig:unfrust} together with the circular magnetic field. As formulated by Owen and Scalapino, we are in the `0 to 1 vortex mode' and the current peak can be thought of as `half' a current vortex present. The Josephson current is limited by the nucleation of a vortex at the top of the cylinder which then moves down, driven by the Lorentz force, and yields a voltage. Alternatively, the critical current corresponds (through the boundary condition) to the effective lower critical field $ H_1 $ for the penetration of vortices into the junction from the top\cite{tinkham:2004},
\begin{equation}\label{eq:crit_unfrust_long}
H_1  = \frac{2 I_c}{c R} = \frac{ 8 \pi }{c} \lambda_J J_c \propto J_c^{1/2}. 
\end{equation}

For short cylinders, $ h \ll \lambda_J $, the current flows nearly uniformly and the critical current is simply, 
\begin{equation}
I_{c} = 2 \pi R h J_c =  I_{c0}.
\label{eq:simple-Ic}
\end{equation}


\subsection{Phase B: frustrated case}

We now turn to phase~B where the situation is very different. Here, the phase winding does not vanish, $N=\pm1$, and the junction is frustrated. What does this mean? It is obvious that for the previous unfrustrated case we can minimize the free energy $f_{SG}$ in Eq.~(\ref{eq:sgfree}) in both the gradient and the coupling part simultaneously at any point on the interface by choosing the phase $ \phi(\theta,z) =  2 \pi n $. However, this is not the case anymore for phase~B. Although keeping the phase constant everywhere still optimizes the gradient part, this yields a coupling part
\begin{equation}
- \frac{1}{\lambda_J^2} \cos \left[\phi  \mp \theta  \right] 
\end{equation}
varying between $ - \lambda_J^{-2} $ and $ + \lambda_J^{-2} $ along $ \theta $. On the other hand, optimizing the coupling term
by choosing $ \phi(\theta,z) = \pm \theta + 2\pi n $ would lead to the problem of a non-vanishing gradient part
\begin{equation}
\frac{1}{2} \left( \frac{\partial \phi}{R\partial \theta } \right)^2 = \pm \frac{1}{2 R^2}.
\end{equation}
This is a typical frustration situation, where the Josephson coupling energy can only be optimized at the expense of magnetic energy and vice versa. 


\subsubsection{Without external current}

We first consider the situation where no external current is applied, $ I=0 $. The junction is then translationally invariant along the $z$-axis and we again have a one-dimensional problem, now independent of the height of the junction, where the phase depends only on $\theta$ and is constant along $z$, $ \phi(\theta,z) = \phi(\theta) $. This case was discussed before in Ref.~\onlinecite{kaneyasu:2010}.

\begin{figure}[htb]
\includegraphics[width=0.95\columnwidth]{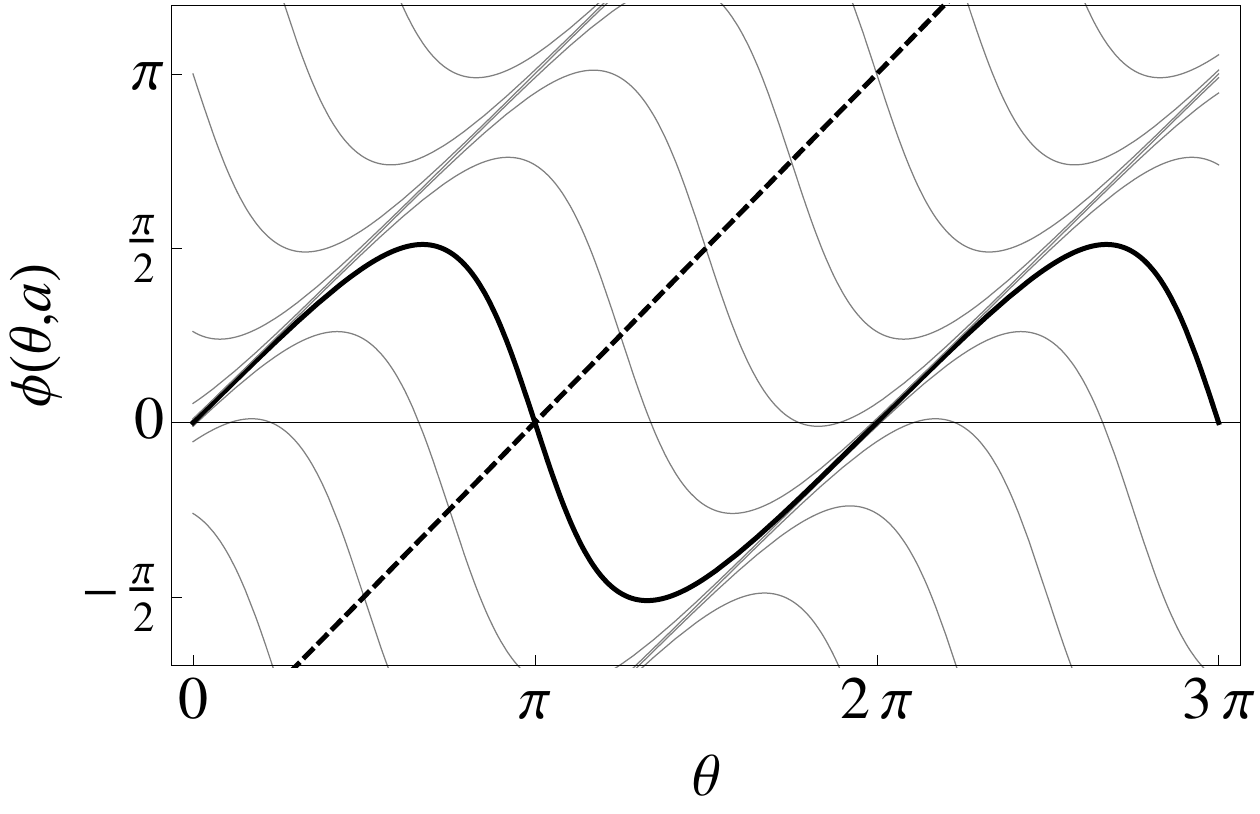}
\caption{\label{fig:phase} Josephson phase in the frustrated case $N=\pm1$ at zero applied current (analytical result). The Josephson penetration depth is $\lambda_J=0.5 R$ and the chirality is positive ($N=+1$). There is a rotational degeneracy and the phase can be shifted along $\theta-\pi$ (dashed) due to the complete rotational symmetry of the junction by any angle without changing the energy. We show multiple solutions for $a=-2\pi+\pi/4\,n$ (integer $n$) with the solution for $a=\pi$ in bold.}
\end{figure}

\begin{figure}[htb]
\includegraphics[width=0.95\columnwidth]{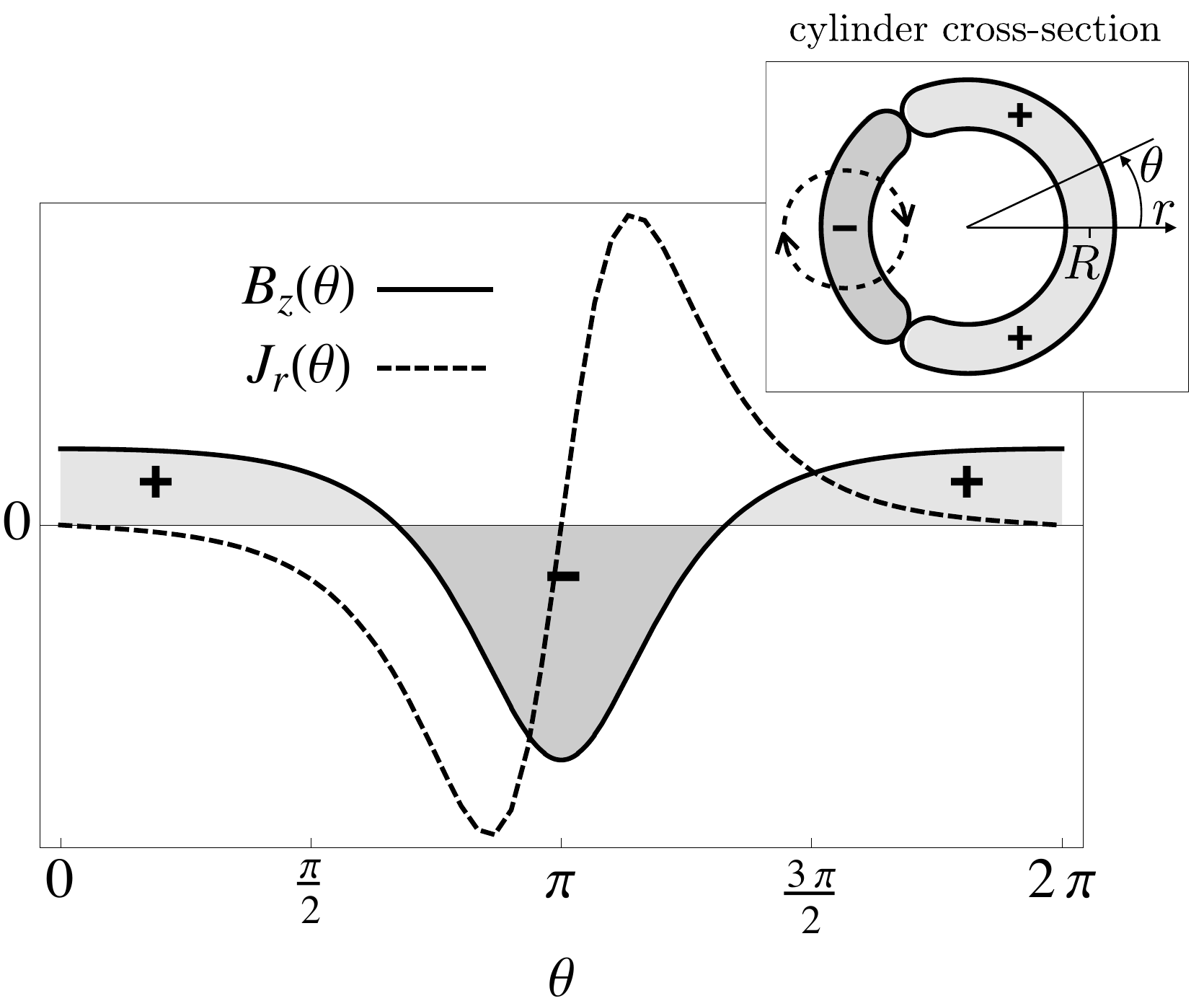}
\caption{\label{fig:frust0} Flux pattern and radial current density in the frustrated case $N=\pm1$ at zero applied current (numerical result). The Josephson penetration depth is $\lambda_J=0.5 R$ and the chirality is positive ($N=+1$). There is a flux line and a current vortex centered at $a=\pi$. The total flux through the cylinder and the net current through the interface are zero. The inset shows the circular cross-section of the cylinder with the sign of the flux and the position of the current vortex indicated.}
\end{figure}

Here, the minimization of the free energy including the periodic boundary condition $ \phi(\theta + 2 \pi)= \phi(\theta) $ is possible analytically and the phase is given by
\begin{equation}
\phi(\theta,a)=\pi-2\text{am}\left(\frac{\text{K}(m)}{\pi}(\theta-a),m\right)+\theta\mod2\pi,
\end{equation}
where $\text{am}(x,m)$ is the Jacobi amplitude function and $\text{K}(m)$ the complete elliptic integral of the first kind. The Jacobi parameter $m\in[0,1)$ is determined via
\begin{equation}
\frac{R}{\lambda_J}=\frac{\sqrt{m}\,\text{K}(m)}{\pi}.
\end{equation}
The parameter $a\in[0,2\pi]$ denotes an undetermined shift as illustrated in Fig.~\ref{fig:phase} where we show the phase for different values of $a$. This degeneracy is because the junction is rotationally symmetric and the phase can be shifted along the identity by any angle without changing the energy.
The magnetic field is finite, and also the current flowing through the interface,
\begin{equation}
\bm{B} =B_z(\theta,a)\hat{\bm z} = \frac{\Phi_0}{2 \pi d } \frac{\partial \phi}{R\partial \theta}=\frac{\Phi_0}{2 \pi d R} \left(1-2\frac{\text{K}(m)}{m}\text{dn}\right)
\end{equation}
and
\begin{align}
\bm{J} &= J_r(\theta,a)  \hat{\bm{r}} = \frac{c}{4 \pi R} \frac{\partial B_z}{\partial \theta} \\
&= \frac{c\Phi_0}{8 \pi^2 d} \frac{\partial^2 \phi}{ R^2\partial \theta^2}= \frac{c\Phi_0}{8 \pi^2 dR^2}\frac{2m\text{K}(m)^2}{\pi^2}\text{cn}\,\text{sn},\nonumber
\end{align}
where cn, dn and sn are the Jacobi elliptic functions with the same argument as the Jacobi amplitude function in the phase above. Here the effect of the degeneracy $a$ is a simple shift around the cylinder. Note that this solution does not resolve the spatial dependence within the range $d$ perpendicular to the interface where screening currents run. Obviously, the total flux vanishes,
\begin{equation}
\int_0^{2 \pi} \!R \text{d}\theta \,B_z(\theta) =   \frac{\Phi_0}{2 \pi d} \left(\phi(2 \pi) - \phi(0)\right) = 0 ,
\end{equation}
and also no net current flows through the interface,
\begin{equation}
\int_0^{2 \pi}\!R \text{d}\theta\, J_r (\theta) =   \frac{c}{4 \pi} (B_z(2 \pi) - B_z(0)) = 0,
\end{equation}
as is expected when no current is supplied from outside. There is a current vortex centered at $\theta=a$ around a negative magnetic flux peak having a width of $ \Delta \theta \propto \lambda_J / R $. The stronger the Josephson coupling, and therefore the smaller $ \lambda_J $, the larger and more concentrated this flux peak becomes. In the extreme limit this would be a flux line enclosing approximately one flux quantum $ - \Phi_0 $ which is compensated by a positive counter flux spread over the remaining part of the interface. In Fig.~\ref{fig:frust0} we show the magnetic flux and current pattern for the state with positive chirality ($N = + 1$) and $a=\pi$. 

\subsubsection{With external current}

We now examine the behavior of the junction if a non-vanishing external current is applied, $ I > 0 $. To analyze this situation we assume that $ \lambda_J \gg R, h $ such that $ \epsilon = R^2 / \lambda_J^2 $ is small while $ h $ and $ R $ are comparable. We then expand the phase obtained from solving the sine-Gordon equation~(\ref{eq:sg}) in $ \epsilon $,
\begin{equation}
\phi(\theta,z) = \phi_0 +  \sum_{k=1}^{\infty} \epsilon^k \phi_k(\theta,z)  ,
\label{eq:eps-expan}
\end{equation}
where $ \phi_0 $ is a constant, as we will show below, and the remaining terms are small perturbations. 
The boundary conditions are
\begin{equation} 
\left. \sum_k \epsilon^k \frac{\partial \phi_k}{\partial z} \right|_{z} = \left\{ \begin{array}{ll}  \gamma & \quad z = h, \\ & \\ 0 & \quad z=0 . \end{array} \right.
\label{eq:bc-eps}
\end{equation}

First, we consider the lowest order only. Inserting the expansion into the differential equation, the coupling part vanishes and we obtain
\begin{equation}
 R^2\nabla^2\phi_0=R^2 \left( \frac{1}{R^2} \frac{\partial^2 \phi_0}{\partial \theta^2} + \frac{\partial^2 \phi_0}{\partial z^2} \right)=0,
\end{equation}
which can only be solved by $ \phi_0 $ being a constant with the boundary condition $ \partial_z \phi_0 |_{z=0,h} = 0 $, since at this order there is no Josephson contact and thus no current can flow. 

Next, we continue with the higher orders, i.e.\ $ k>0 $. By expansion in $ \epsilon $ we obtain differential equations of the Poisson form
\begin{equation}
\nabla^2 \phi_k (\theta,z) = q_k (\theta ,z).
\end{equation}
The source term $ q_k (\theta ,z) $ does not depend on $ \phi_k(\theta,z) $, nor on $ \epsilon $, but it does depend on the resulting $ \phi_{k'}(\theta,z) $ from lower orders $k'<k$ (iterative approach). We also obtain Neumann boundary conditions at the edges of the cylinder
\begin{equation}
\left. \frac{\partial \phi_k }{\partial z} \right|_{z} = \left\{ \begin{array}{ll} c_k  & \quad z=h , \\ & \\ 0 & \quad z=0 ,\end{array} \right.
\end{equation}
which combine, together with Eq.~(\ref{eq:bc-eps}), to the condition
\begin{equation}
\gamma = \sum_{k=0}^{\infty} \epsilon^k c_k .
\end{equation} 
The periodic boundary conditions in $\theta$ still apply,
\begin{equation}
\phi_k (\theta+2 \pi,z) = \phi_k (\theta,z).
\end{equation}
For a solution to exist, the compatibility condition for the Neumann problem requires that
\begin{equation}
c_k  = \frac{1}{2\pi R}\int_{0}^{2 \pi}\!R \text{d}\theta \int_0^h\!\text{d}z \, q_k (\theta, z) .
\end{equation}

Let us now consider the actual equations order by order. We start with the first order in $ \epsilon $,
\begin{equation} 
R^2 \nabla^2 \phi_1 = \sin (\phi_0 - \theta),
\label{eq-hier-1}
\end{equation} 
which is obviously only compatible with $ c_1 = 0 $ since $\phi_0$ is a constant. The solution is then given by
\begin{equation}
\phi_1(\theta,z) = - \sin (\phi_0 - \theta),
\end{equation}
where any integration constant can be absorbed in the constant $ \phi_0 $ as part of the complete solution $\phi$. Using this solution in turn for the equation of next order in $ \epsilon $, we obtain,
\begin{equation}
R^2 \nabla^2 \phi_2  = \phi_1(\theta,z) \cos(\phi_0 - \theta) =   - \sin(\phi_0 - \theta)  \cos(\phi_0 - \theta)
\end{equation}
which is again only compatible with $ c_2 = 0 $. The solution is now given by
\begin{equation}
\phi_2(\theta, z) =  \frac{1}{4}   \sin(\phi_0 - \theta)  \cos(\phi_0 - \theta) .
\end{equation}
This solution can again be used to derive the next order in a hierarchy of Poisson differential equations that are all compatible with the Neumann boundary conditions only if $ c_k =0 $ for all orders $k$. 
Therefore, no solution exists for a finite value of $ \gamma $. 

This result is supported by our numerical analysis which also does not yield a stable solution. Rather, we obtain solutions varying in time. Referring to the resistively shunted junction model, this is due to a dissipative contribution from a junction resistance $ R_J $ besides the Josephson channel \cite{orlando:1991}.
We take this effect into account by an additional term in the sine-Gordon equation~(\ref{eq:sg}), 
\begin{equation}
-\tau \frac{\partial \phi}{\partial t}+ \frac{\partial^2\phi}{R^2\partial\theta^2}+\frac{\partial^2\phi}{\partial z^2}=\frac{1}{\lambda_J^2}\sin(\phi-\theta) ,
\label{eq:sg-t}
\end{equation}
where
\begin{equation} 
\tau = \frac{2 d}{c R h R_J}  .
\end{equation}
Assuming that the externally applied current is small, we approximate the solution in zeroth order in $\epsilon$ as
\begin{equation}
\phi_0(\theta, z, t) = -\frac{\gamma}{h}\frac{t}{\tau} + \frac{\gamma}{2} z^2 + \tilde{\phi}_0,
\end{equation}
where $ \tilde{\phi}_0 $ is a constant. The boundary condition for a finite current is then already satisfied by $ \phi_0 $, and we approximate all the other $ \phi_k $ with $ k \geq 1 $ by ignoring the $z$-dependence and using $ c_k = 0 $. The (constant) voltage is given by
\begin{equation}
V = \frac{\Phi_0}{2 \pi}   \frac{\partial \phi}{\partial t} = - \frac{\Phi_0}{2 \pi} \frac{\gamma}{\tau h} = R_J I,
\end{equation}
which corresponds to Ohmic behavior. The temporal dependence of the higher order components $ \phi_k $ is oscillatory and ignored here by taking a time average. This rather simple approximation of the extended model already incorporates the basic observation from the numerical solution of the full problem that the flux pattern moves around the cylinder, as the phase $\phi$ is essentially linear in $t$. Thus, we conclude that there exists a solution of the sine-Gordon equation for any finite current, but with a time-dependent phase $\phi$ corresponding to a resistive current flow.

Thus, the homogeneous frustrated junction does not support any finite supercurrent, since the position of the magnetic flux and current pattern induced by the frustration is not fixed. Rather it can be shifted by any angle without changing the energy. Thus, any finite current will drive the flux pattern by the Lorentz force. Dissipative dynamics then lead to a steady state situation with a constant time-averaged voltage. This will obviously also happen in the other limit, $ \lambda_J \ll  R, h $.


\subsection{Phase B: inhomogeneous junction}

We now change the situation by removing the rotational symmetry which is the underlying reason why the frustrated junction does not support any supercurrent. Maintaining the cylindrical geometry within our model, this can be implemented by introducing an angular dependent Josephson coupling $ \tilde{J}_c (\theta) $ by adding a modulation
\begin{equation}
\frac{1}{\lambda_J^2} \longrightarrow \frac{1}{\lambda_J^2} \{ 1 + m(\theta) \} 
\label{eq:aniso-jc}
\end{equation}
with the restrictions that $ m(\theta) > -1 $, $ m(\theta + 2 \pi) = m(\theta) $, and
\begin{equation}
\int_0^{2 \pi}\!\text{d}\theta\, m(\theta) = 0  ,
\end{equation}
whereby $ J_c $ of Eq.~(\ref{Jc-lambda}) shall be the angle-averaged coupling constant.
The differential equation now reads
\begin{equation}
R^2 \nabla^2 \phi = \epsilon \{ 1 + m(\theta) \}  \sin(\phi - \theta),
\end{equation}
while the boundary conditions remain the same.

Again considering $ \epsilon $ small we use the expansion from Eq.~(\ref{eq:eps-expan}) to analyze the effect of the variable $\lambda_J$.
For $ \phi_0 $ we still obtain a constant while the equation for the first order in $\epsilon$ is now
\begin{equation}
R^2\nabla^2\phi_1=\frac{\partial^2 \phi_1}{\partial \theta^2} + R^2 \frac{\partial^2 \phi_1}{\partial z^2} =  \{ 1 + m(\theta) \} \sin (\phi_0 - \theta)  .
\end{equation}
The compatibility condition here leads to
\begin{equation}
c_1= \frac{h}{R^2}\tilde{m}(\phi_0),
\end{equation}
where
\begin{align}
\tilde{m}(\phi_0)  &=\int_{0}^{2 \pi}\! \frac{\text{d}\theta}{2\pi}  \,  m(\theta)  \sin (\phi_0 - \theta) \nonumber\\
&= \left(m_1 \sin \phi_0 - m_2 \cos \phi_0\right) 
\end{align}
with
\begin{equation} 
\hat{m} = m_1 + i m_2 =  \int_0^{2 \pi}\! \frac{\text{d}\theta}{2 \pi}\, m (\theta) e^{i \theta}.
\end{equation}
We now find the complete solution
\begin{equation}
\phi_1(\theta ,z) = \varphi (\theta, \phi_0)+\frac{\tilde{m}(\phi_0)}{2 R^2} z^2
\end{equation}
fulfilling the boundary conditions, and where $ \varphi(\theta,\phi_0) $ is the special solution of the differential equation
\begin{equation}
\frac{\partial^2 \varphi}{\partial \theta^2} = \{ 1 + m(\theta)  \} \sin(\phi_0 - \theta) - \tilde{m}(\phi_0)  .
\end{equation}

It is obvious that we can satisfy the boundary conditions for finite currents $ I \neq 0 $ as long as $ \tilde{m}(\phi_0) $ is not zero. Neglecting higher orders, the compatibility condition now leads to
\begin{equation}
\gamma =  \epsilon \left. \frac{\partial \phi_1}{\partial z} \right|_{z=h} = \frac{h}{\lambda_J^2} \tilde{m}(\phi_0) 
\end{equation}
which corresponds to the current-phase relation
\begin{equation}
I(\phi_0) = 2 \pi R h J_c \tilde{m}(\phi_0) = I_{c0}  \tilde{m}(\phi_0) ,
\end{equation}
where $ I_{c0} $ is given by Eq.~(\ref{eq:simple-Ic}) as the integral of the critical current density $ J_c $ over the interface. With $ \left|\tilde{m}(\phi_0)\right| < 1 $ for all $ \phi_0 $ this current $I(\phi_0)$ is always smaller than $I_{c0}$ and its maximum, the renormalized critical current, is given by
\begin{equation}
I_c = I_{c0} \max_{\phi_0} \tilde{m}(\phi_0) = I_{c0} \sqrt {m_1^2 + m_2^2} = I_{c0} \left| \hat{m}\right|  < I_{c0}  .
\end{equation}

Strictly speaking we have to take into account the variable Josephson coupling also in the unfrustrated case used as a reference. However, in the limit $\lambda_J\gg R,h$ the junction is `short' in both directions and any effects on length scales shorter than the Josephson penetration depth are averaged out. The resulting current density simply adjusts to the local critical current density like $J_r(z,\theta)=J_{r,0}(z)(1+m(\theta))$ with $J_{r,0}$ the result of the homogeneous case. Thus, the critical current of the inhomogeneous unfrustrated case is the same as for the corresponding homogeneous case with the same average Josephson coupling. This is confirmed both by numerics and a similar analytical analysis as above. Also, it is in accordance with previous results as summarized by Barone and Patern\`o in Sec.~4.4. of Ref.~\onlinecite{barone:1982}.

\begin{figure}[htb]
\includegraphics[width=0.95\columnwidth]{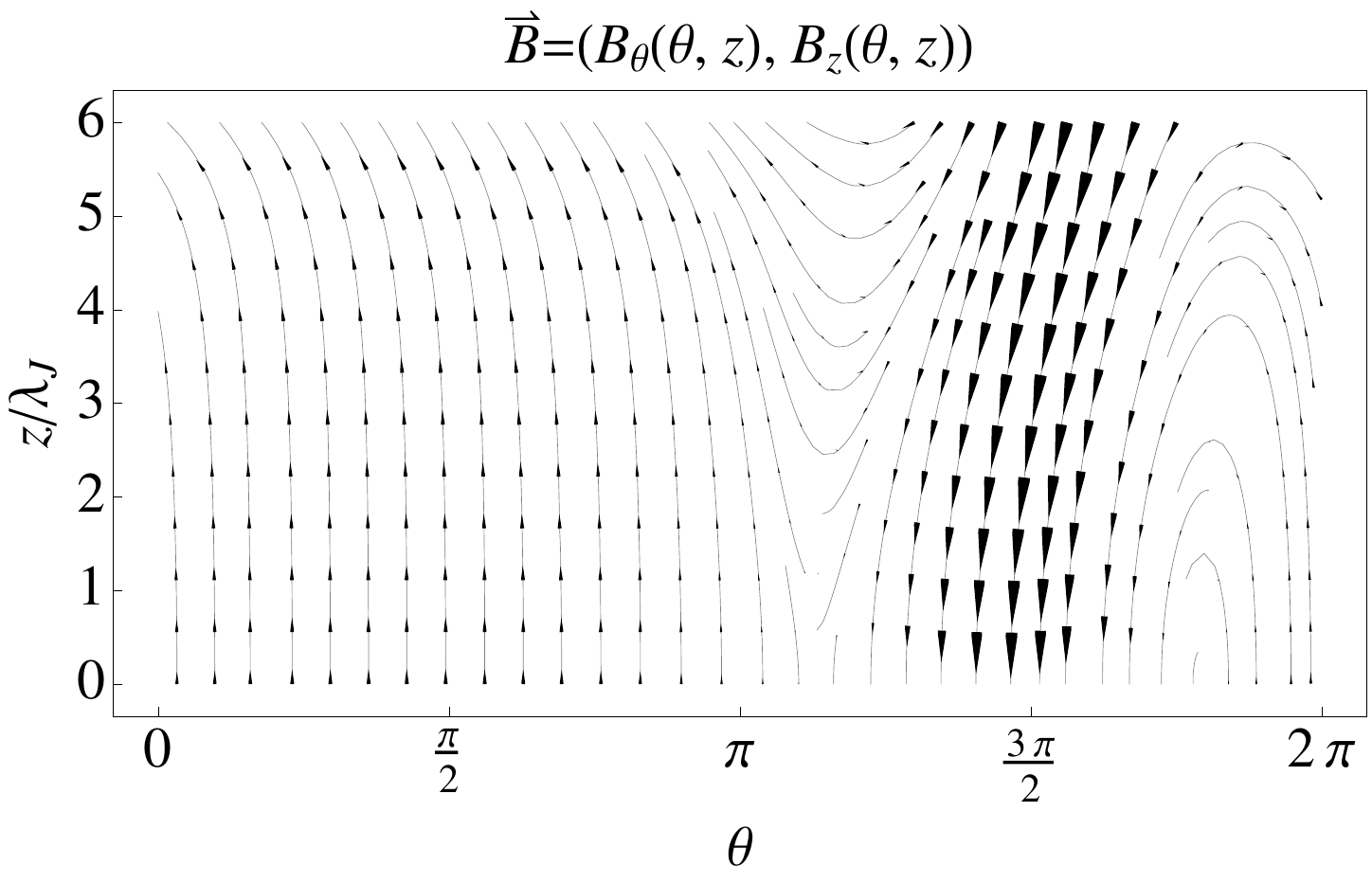}
\caption{\label{fig:frust} Deformed flux pattern in the frustrated case $N=\pm1$ for an inhomogeneous junction at the renormalized critical current $I=I_{c}$. The Josephson penetration depth is $\lambda_J=0.5 R$ and the chirality is positive ($N=+1$). The height of the cylinder is $h=3 R$ and in terms of the Josephson penetration depth it is $h=6\lambda_J$. The inhomogeneity is created by introducing an angular dependent Josephson coupling implemented by modulating the coupling strength with $m(\theta)=0.2\cos(\theta)$. The flux line appearing spontaneously already at zero applied current has a fixed initial position at the minimum of the modulation, here at $\pi$, where it is pinned by the junction inhomogeneity. At a finite applied current it is shifted away in the direction of the Lorentz force (here counter-clockwise). Due to screening the externally induced current flows through the interface mainly at the top. Thus, the flux line is bent near the top where the Lorentz force is strongest.}
\end{figure}

We conclude that the inhomogeneous frustrated junction does support a finite supercurrent up to the renormalized critical current. Considering the junction dynamics, the mechanism involved here is based on the Lorentz force effect. As the applied current is increased, the magnetic flux and current pattern is shifted away from its stable initial (no applied current) position which is now fixed because of the angular dependent Josephson coupling. In this process, the basic shape of the flux peak is preserved while its orientation along the $z$-axis is no longer straight. Since the current is screened inside the junction (in analogy with the unfrustrated case), the phase now also depends on $z$ and the flux line is shifted farther away from its initial position near the top edge of the cylinder where the current is injected and the Lorentz force therefore is the strongest. This deformation is illustrated in Fig.~\ref{fig:frust} where we show the magnetic field $\boldsymbol{B}=B_\theta \hat \theta+B_z\hat z$ on the interface of the junction with the flux peak compensated by a weak but broad counter flux. Even at a finite applied current, the flux line pattern does not move because it is pinned by the junction inhomogeneity.

Once the applied current is larger than the renormalized critical current, the resulting Lorentz force is strong enough to overcome this pinning potential. This corresponds to a depinning transition of the magnetic flux and current pattern which then starts to move, resulting in dissipation with a non-vanishing voltage in the same way as discussed above. Naturally, this pinning effect also governs the dynamics in the limit $ \lambda_J \ll  R, h $, where it is even stronger, as we will discuss below.

The behavior discussed within this approximative approach is confirmed by the full numerical solution of the differential equation above. Using a relaxation process (diagonally preconditioned quasi-Newton method for the free energy) to solve the boundary value problem, it is rather easy to test whether a solution exists or not, as the iteration either converges (pinned flux pattern) or runs without convergence (depinned flux pattern) again leading to a solution varying in time as discussed above. 

We have found here a limiting mechanism for the Josephson current for the inhomogeneous frustrated junction with a renormalized critical current always below the critical current from the unfrustrated case. This can also be understood from the fact that the flux pattern involved in the dynamics here is spontaneously present within the junction even without an applied current while in the unfrustrated case they first have to be nucleated at the boundary of the junction.

\section{Discussion of the pinning effect}\label{sec:pin-strong}

In this section we discuss in detail the pinning effect appearing in the inhomogeneous frustrated junction, focusing on the difference between the two limits of weak $ \lambda_J \gg R,h $ (the approximative approach outlined above), and strong coupling $ \lambda_J \ll R,h $ (a well-localized flux line with broad counter flux). We give analytical arguments for the behavior of these limits, while the interpolation between them is possible numerically.


\subsection{Pinning energy}

The first quantity we address is the pinning energy,  a measure of the depth of the pinning potential associated with a a given structure of an inhomogeneous Josephson coupling at zero applied current. Defining the coupling anisotropy again as in Eq.~(\ref{eq:aniso-jc}), we label the free energy density $ f_{SG}[m(\theta)] $ through the presence of a modulation $ m(\theta) $. We then define the pinning energy as 
\begin{equation}
E_\text{pin}  =  \int_0^{2 \pi} \!R \text{d} \theta\int_0^h\!\text{d}z\, \left\{ f_{SG}[m(\theta)=0] - f_{SG}[m(\theta)] \right\}.
\end{equation}

We first consider the weak coupling limit where the above solutions can be used. No current $ I=0 $ means $ \tilde{m}(\phi_0)= 0 $. To linear order in $ \epsilon $ we then find
\begin{equation}
E_\text{pin} \approx \frac{2 \pi R h}{ \lambda_J^{2}} | \hat{m} |  \propto J_{c}  .
\end{equation}

Turning next to the strong coupling limit we have to consider a different solution of the sine-Gordon equation~(\ref{eq:sg}). As shown in Ref.~\onlinecite{kaneyasu:2010}, a good approximation for the phase $\phi$ in a homogeneous junction with no applied current is given by the soliton solution
\begin{equation}
\phi(\theta,z) = \theta  - 4 \arctan \left(e^{(\theta - u) \Lambda} \right),
\label{eq:soliton}
\end{equation}
where $ \Lambda = R/\lambda_J \gg 1 $. The parameter $ u$ determines the (variable) position of the flux line along $\theta$, which for no applied current is also the center of the spontaneous current vortex. The soliton is localized enough to satisfy the periodic boundary condition in $\theta$. For the inhomogeneous junction case we assume the same basic soliton shape but with the initial position $u_0$ being fixed by the presence of the coupling anisotropy. It is then straightforward to calculate the pinning energy for this case as
\begin{equation} 
E_\text{pin} \approx \frac{4h}{\lambda_J} \left(-m(u_0)\right) \propto \sqrt{J_c}  ,
\end{equation}
where $ u_0 $ is the position of the minimum of $ m(\theta) $. As expected, the spontaneous current vortex is centered on the minimum of the pinning potential where the Josephson coupling strength is the weakest.

\begin{figure}[htb]
\includegraphics[width=0.9\columnwidth]{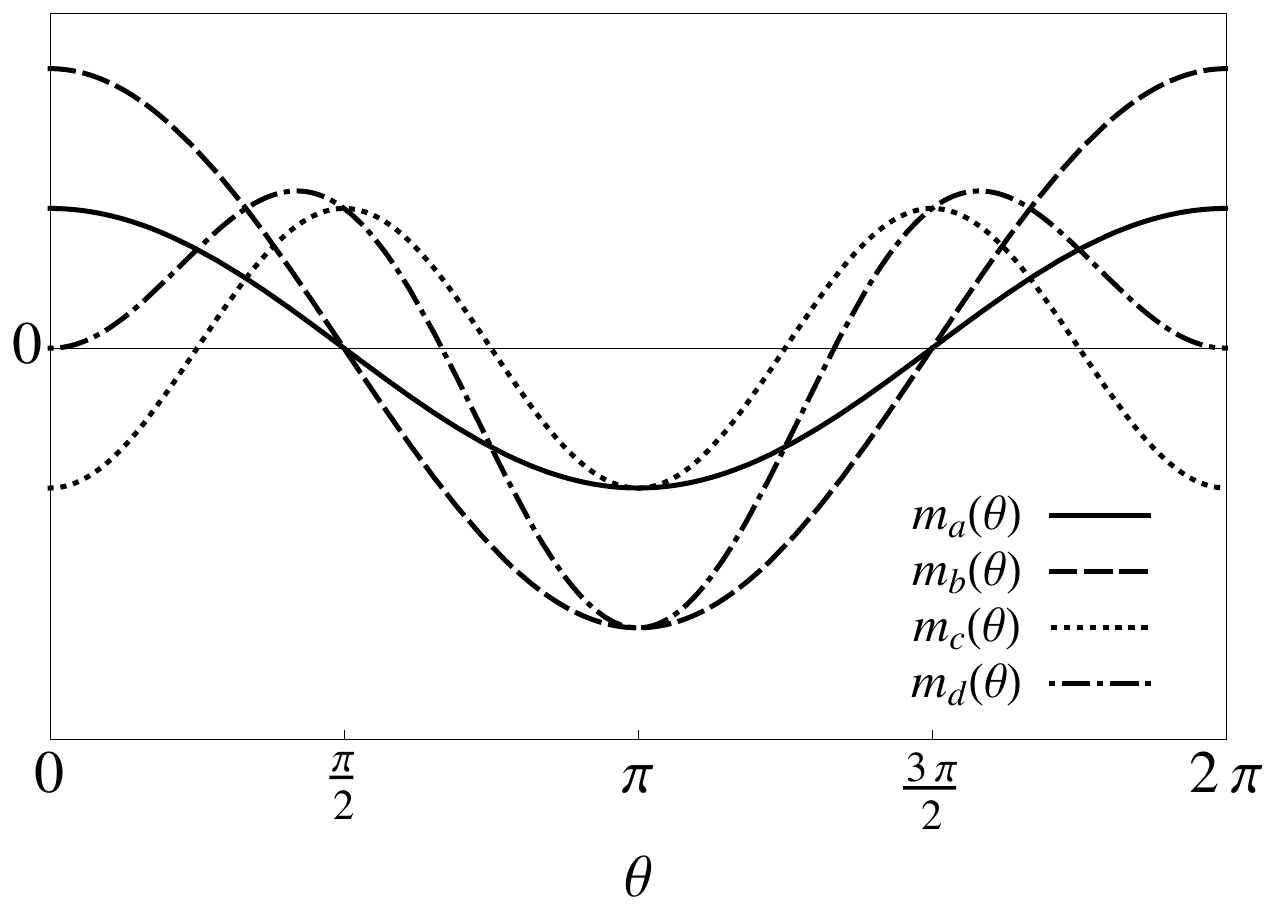}
\caption{\label{fig:mod} The four sample modulations $\textrm{m}(\theta)$. We consider cosines with different amplitudes, a second order cosine, and a combination: ${m_a=0.25\cos(\theta)}$, ${m_b=2m_a=0.5\cos(\theta)}$, ${m_c=-0.25\cos(2\theta)}$, and ${m_d=m_a+m_c}$. As required, the integral over one period is zero. The renormalization factor $|\hat{m}|$ is given by half the first cosine coefficient and is the same for $m_a$ and $m_d$, while it vanishes for $m_c$. The minimum and thus the initial position is always $u_0=\pi$. The minimum value $-m(\pi)$ is the same for $m_a$ and $m_c$, and for $m_b$ and $m_d$. The amplitude $\Delta m$ is the same for $m_a$ and $m_c$.}
\end{figure}

\begin{figure}[htb]
\includegraphics[width=0.95\columnwidth]{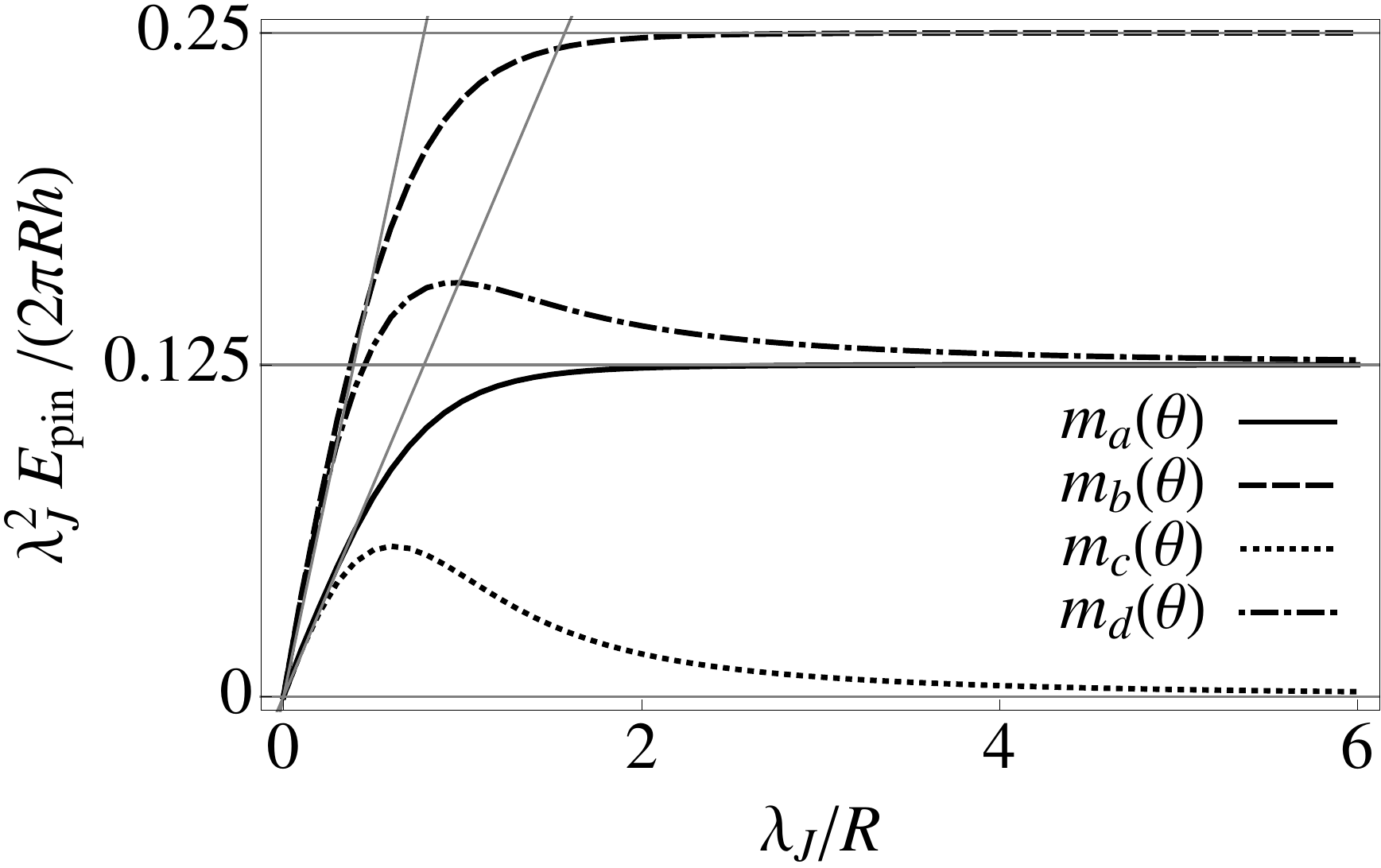}
\caption{\label{fig:pin} Numerical results of the pinning energy for the four sample modulations in Fig.~\ref{fig:mod}. We show $\lambda_J^2E_{pin}$ in units of $2\pi Rh$ against $\lambda_J$ and also indicate the limiting behavior.
In the weak coupling limit $\lambda_J\gg R,h$, the pinning energy saturates to a constant value given (in these units) by the renormalization factor $|\hat{m}|$, which is half the first cosine coefficient, and is 0.125 for $m_a$ and $m_c$, 0.25 for $m_d$, and vanishes for $m_b$.
In the strong coupling limit $\lambda_J\ll R,h$, the pinning energy increases linearly with the slope proportional to the minimum of the modulation which is the same for $m_a$ and $m_c$, and for $m_b$ and $m_d$.
}
\end{figure}

These limiting behaviors are well reproduced in our numerical treatment as illustrated in Fig.~\ref{fig:pin} where we show the full numerical results for the pinning energy for different sample modulations and specifically indicate the limiting behavior found through the analytical considerations above. We use different sample modulations $m(\theta)$ shown in Fig.~\ref{fig:mod} based on cosines of different amplitudes and orders. There is an obvious regime change at $ \lambda_J \sim R $. In summary we find
\begin{equation}
 \lambda_J^2E_\text{pin} \propto \left\{ \begin{array}{ll} \lambda_J &\quad \lambda_J \ll R,h, \\ & \\ \text{const.} &\quad \lambda_J \gg R,h. \end{array} \right. 
\end{equation}

\subsection{Critical current}
 
Next, we investigate the critical current itself, again considering the two limits. The same type of scaling we found for $ E_\text{pin} $ is also visible here.
The weak coupling limit has already been discussed above and we found
\begin{equation}
I_c  \approx \frac{c\Phi_0Rh}{4 \pi d} \frac{| \hat{m}|}{\lambda_J^2 }=I_{c0}| \hat{m}|\propto J_c
\end{equation}
which depends directly on the modulation. 

We then turn to the strong coupling limit. When calculating the pinning energy we found that the initial position of the phase soliton from Eq.~(\ref{eq:soliton}) is at the point of the weakest coupling, i.e.\ the minimum of $ m(\theta) $. To analyze this limit further we now suggest treating the pinned magnetic flux line as an elastic string in a potential landscape. With a non-vanishing external current applied, this string will deform while the basic soliton shape is preserved. This leads to the simple approximation,
\begin{equation}
\phi(\theta,z) = \theta + v(z) - 4 \arctan\left(e^{[\theta - u(z)] \Lambda }\right),
\label{eq:ansatz}
\end{equation}
where $ u(z) $ is the displacement of the flux line in the $ \theta$-direction and $ v(z) $ a $z$-dependent phase shift.
Based on the solution of the unfrustrated case, and confirmed below by the effective approach described in the Appendix, we approximate $v(z) $ as
\begin{equation}
v(z) = \gamma \lambda_J e^{(z-h)/\lambda_J} ,
\label{eq:unfrust_sol}
\end{equation}
which already fulfills the boundary conditions for the applied current in Eq.~(\ref{bc-phi}). The boundary condition for the displacement $u(z)$ is therefore $ \partial_z u(z=h) = 0 $.

Note that $ v(z) $ has an influence only very close to $ z = h $ within a range of $ \lambda_J \ll h $. Below, we may therefore use an approach with an effective boundary condition for $u(z)$,
\begin{equation}
u'(h) = \frac{\pi \lambda_J}{4 R} \gamma,
\end{equation}
and neglect the influence of $v(z)$. A detailed discussion of this effective model can be found in the Appendix.

We then insert the ansatz from Eq.~(\ref{eq:ansatz}) into the free energy and by making the approximations mentioned above we obtain the following functional for $ u(z) $,
\begin{equation}
F^*[u] = \int_0^h dz \left[ \frac{1}{2} (u'(z) )^2 + \frac{1}{2 R^2} m(u(z))  \right].
\end{equation}
From this we find the variational equation
\begin{equation}
u'' - \frac{1}{2 R^2} m'(u) = 0.
\label{eq:eom}
\end{equation}
This can also be viewed as the equation of motion for a particle whose position is given by a single coordinate $ u $ in a potential $ V(u)= -m(u)/2 R^2 $ with $ z$ playing the role of time. The boundary conditions then correspond to the velocities at the `times' $ z=0 $ and $ h $. The stable initial position at $ u = u_0 $ is a maximum of the potential $V(u) $. For non-vanishing $ \gamma $ and $ h \gg \ell $ (length scale defined in Eq.~(\ref{eq:ell})) we then interpret the situation as follows. At $z=0$ the particle `starts' at $ u(z=0) \approx u_0 $ corresponding to the top of the potential. At the `time' $z=h$ the particle ends up at a position $ u(h)= u_1 $ of lower potential energy and has gained the `kinetic energy'
\begin{align}
 \frac{(u'(h))^2}{2} &  = \left(\frac{\pi \lambda_J \gamma}{4 \sqrt{2} R} \right)^2 = V(u_0) - V(u_1) \nonumber \\ 
 &   = \frac{m(u_1) - m(u_0)}{2R^2}  .
\end{align}
The maximum of this `kinetic energy' corresponds to the maximum possible gain of potential energy,
\begin{equation}
\Delta m  = \max_{u_1} \left\{ m(u_1) - m(u_0) \right\} ,
\end{equation}
such that the maximal value of $ \gamma $ for which there is a solution of the variational equation (\ref{eq:eom}) is given by
\begin{equation}
\gamma_c = \frac{4}{\pi} \frac{\sqrt{\Delta m}}{\lambda_J}  .
\end{equation}
The critical current is therefore
\begin{equation}
I_c = 8  R\lambda_J J_c  \sqrt{\Delta m}\propto \sqrt{J_c},
\end{equation}
which scales similarly as the unfrustrated junction in this limit (see Eq.~(\ref{eq:crit_unfrust_long})),
\begin{equation}
I_{c} = 4\pi R \lambda_J J_c.
\end{equation}
Before comparing the different results we again have to consider how the modulation affects the basic unfrustrated case. In the limit $\lambda_J\ll R,h$ the junction is `long' in both directions and not only the local critical current density $J_c$ changes but the Josephson penetration depth $\lambda_J$ is modified as well. Where $J_c$ is locally enhanced, $\lambda_J$ becomes shorter. Since both these values influence the current density, there is not a simple current redistribution any more and the critical current is lower than in the basic homogeneous case since the region of weaker coupling strength (now on a length scale larger than the Josephson penetration depth) can provide an entry point for a vortex at the top. This is confirmed by numerics. Since also in the frustrated case the penetration of Josephson vortices from above provides the ultimate limit, there exist situations for the inhomogeneous junction where the critical current from the unfrustrated case lies below the theoretical critical current for the frustrated case. Then, the effective critical current for the frustrated case would rather be the one from the standard mechanism. This, again, is confirmed by numerics. However, such situations only exist for very long junctions with a very strong and broad modulation.

In summary we find the same scaling behavior for the critical current at a fixed height as for the pinning energy
\begin{equation}
\lambda_J^2I_c \propto \left\{ \begin{array}{ll} \lambda_J &\quad \lambda_J \ll R,h, \\ & \\ \text{const.} &\quad \lambda_J \gg R,h. \end{array} \right.
\end{equation}

\begin{figure}[htb]
\includegraphics[width=0.95\columnwidth]{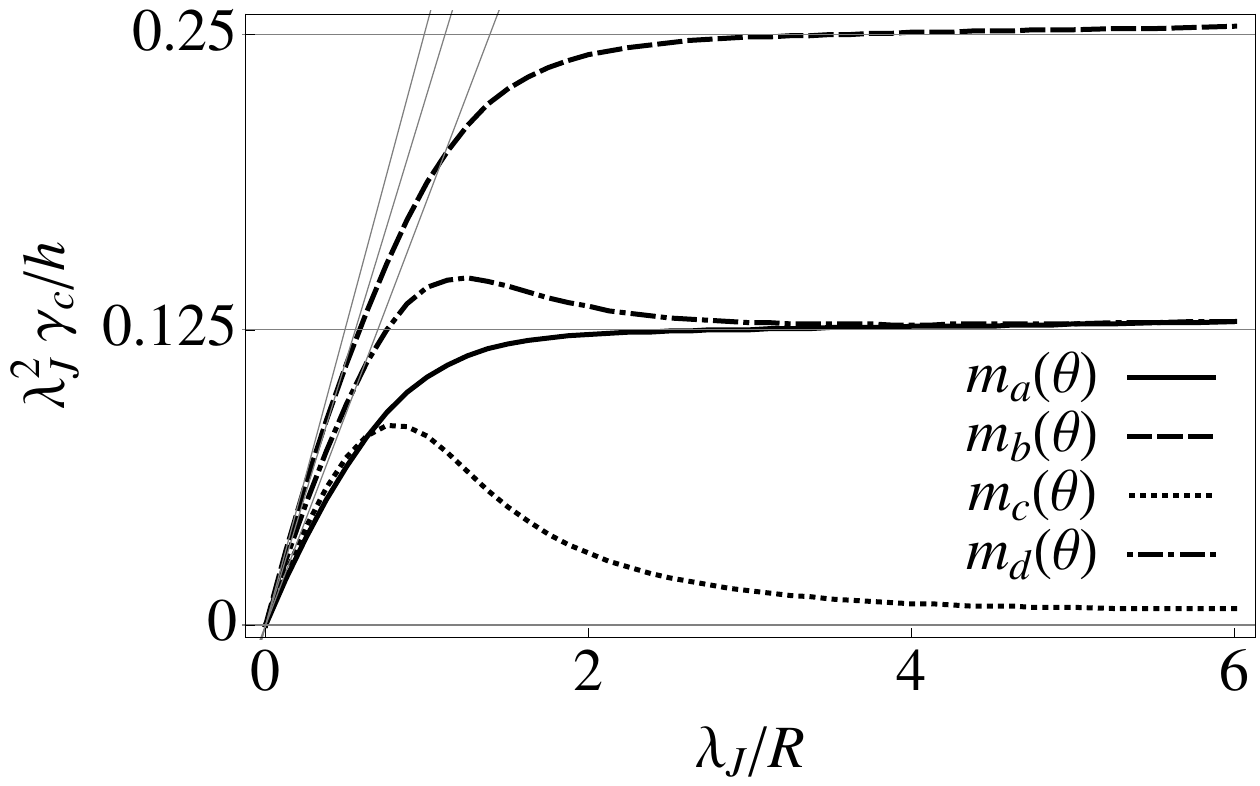}
\caption{\label{fig:crit} 
Numerical results of the critical current at a fixed height $h=5$ for the four sample modulations. We show $\lambda_J^2 I_c/h$ in units of $\frac{c\Phi_0 R}{4\pi d}$ against $\lambda_J$ and also indicate the limiting behavior. The critical current basically behaves like the pinning energy. 
In the weak coupling limit $\lambda_J\gg R,h$, the critical current saturates to a constant value again given (in these units) by the renormalization factor $|\hat{m}|$, which is half the first cosine coefficient, and is 0.125 for $m_a$ and $m_c$, 0.25 for $m_d$, and vanishes for $m_b$.
In the strong coupling limit $\lambda_J\ll R,h$, the critical current increases linearly with the slope now proportional to the square root of the amplitude of the modulation which is only the same for $m_a$ and $m_c$.
}
\end{figure}

As before these limiting behaviors are well reproduced in our numerical treatment. We use the same sample modulations $m(\theta)$ as before, shown in Fig.~\ref{fig:mod}. The scaling behavior at a fixed height $h$ is shown in Fig.~\ref{fig:crit} where we again also indicate the limiting behavior.

\section{Conclusion}\label{sec5}

Recent experiments have studied the Josephson effect between a conventional superconducting Pb film and the presumably chiral $p$-wave superconducting bulk of Sr$_2$RuO$_4$ through Ru-metal inclusions. Maeno~\textit{et al.} found an anomalous temperature dependence of the critical current. In this paper we analyzed a model situation of a single cylindrical Ru-inclusion carrying a superconducting $s$-wave order parameter which is coupled through the interface to the $p$-wave order parameter in the bulk material Sr$_2$RuO$_4$. We have considered two topologies for the $p$-wave order parameter, the trivial and the chiral one, which appear in the 3-Kelvin phase and the bulk phase of Sr$_2$RuO$_4$, respectively. Due to the cylindrical geometry the chiral topology leads to a frustrated junction, while the trivial one is unfrustrated. This frustration causes a rather distinct behavior when an externally applied current runs through the junction. 

While the unfrustrated junction behaves as expected for an extended Josephson junction in both the long and the short junction limit, the frustrated junction is characterized by the appearance of a spontaneous magnetic flux on the interface already at zero applied current. In the case of a rotationally symmetric junction this prevents the flow of a supercurrent since the flux pattern, driven by the Lorentz force, immediately starts to move, leading to dissipation. Only pinning the flux pattern through an inhomogeneous Josephson coupling and thus breaking the rotational symmetry stabilizes the supercurrent, now limited by a pinning-depinning transition. As long as moderate modulations are considered, the critical current is always smaller in the frustrated junction, even if the average coupling strength is the same as in the corresponding unfrustrated junction for which the vortices involved in the dynamics are only created with increasing current and are not spontaneously present.

We therefore propose that this is a possible way to explain the rather sharp drop in the critical current seen at the transition temperature $ T_c \approx 1.5 $~K, associated with a change in the topology of the $p$-wave order parameter. Thus, $ T_c $ sets the boundary between the frustrated and unfrustrated Josephson junction and also between the different topologies for the $p$-wave order parameter of the Sr$_2$RuO$_4$, as will be discussed elsewhere. 

In our model we have simplified the modulation of the Josephson coupling by allowing only angular modulations and keeping it constant along the $z$-axis. This clearly leads to stronger pinning, analogous to columnar pinning for vortex lattices. As a further simplification the spatial dependence of the proximity induced superconducting order parameter in the Ru inclusion is ignored. However, we do not expect qualitative differences concerning the reduction of the critical current in the frustrated case, while the scaling behavior in the two junction limits may be different. Also, there exist extreme situations of very long junctions with a very strong modulation where the frustrated case is effectively limited by the standard mechanism.

Having established a new limiting mechanism for the critical current, the next step is a discussion of the succession of the different states of the junction also taking into account the spatial dependence both along the $z$-axis (proximity effect) and perpendicular to the interface (filamentary nucleation of superconductivity in the {3-K} phase), and also considering domain walls in the bulk Sr$_2$RuO$_4$ material. In future study\footnote{H.~Kaneyasu, S.~B.~Etter and M.~Sigrist, in preparation.} we will present a phase diagram for the whole Pb/Ru/Sr$_2$RuO$_4$ junction further explaining in more detail the anomalous temperature dependence of the critical current.


\begin{acknowledgments}
We are grateful for many valuable discussions with T.~Bzdusek, R.~Willa, D.~Geshkenbein, Y.~Maeno, T.~Nakamura, and the late N.~Hayashi. This study has been financed by a grant of the Swiss National Science Foundation. Moreover, H.~K. is grateful for financial support from the Japan Securities Scholarship Foundation and for the hospitality of the Pauli Center for Theoretical Studies during her visits at ETH Zurich. \end{acknowledgments}


\appendix
\section{Effective strong coupling model}\label{sec:app}

\begin{figure}[htb]
\includegraphics[width=1\columnwidth]{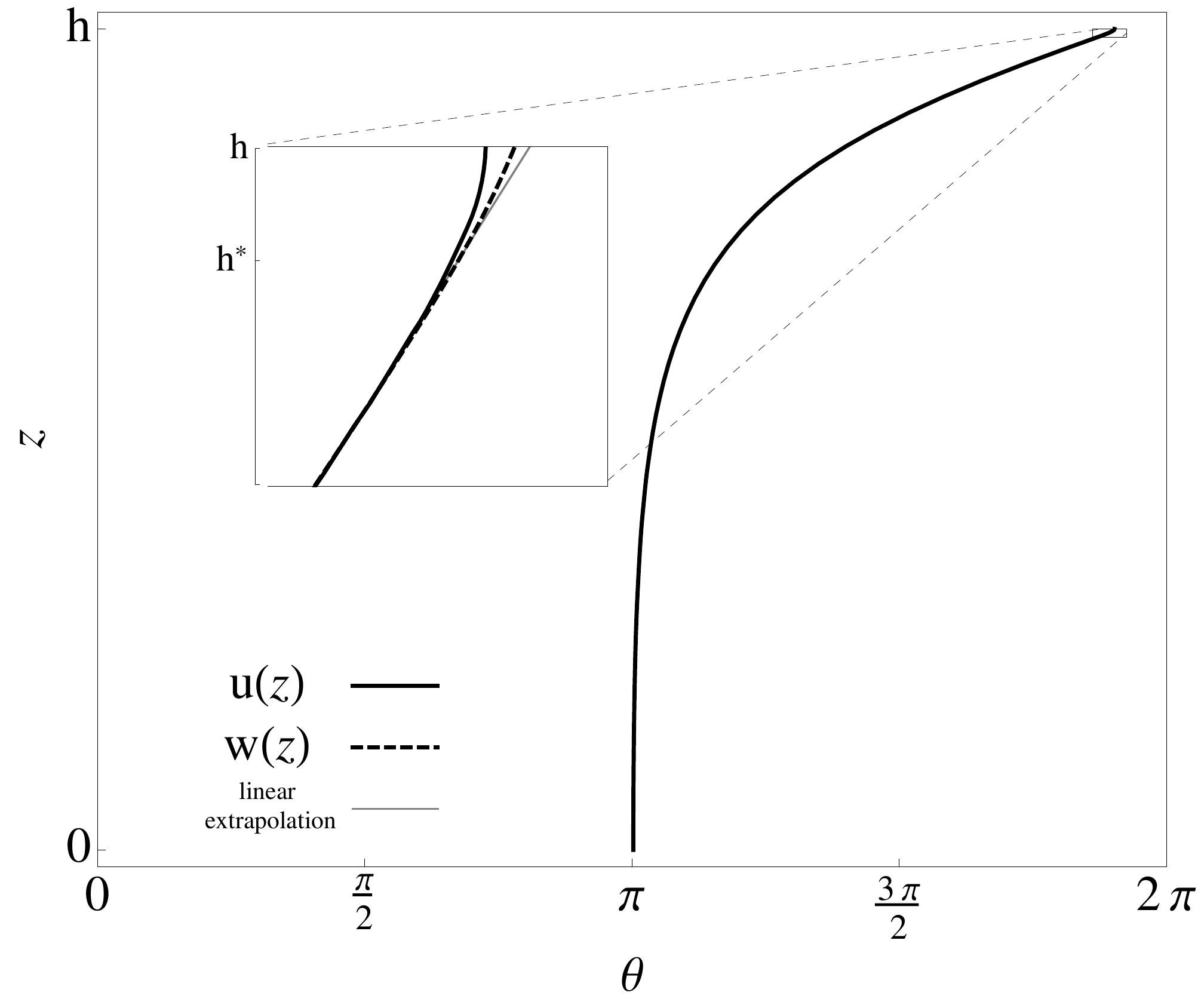}
\caption{\label{fig:vort} Flux line displacement $u(z)$ and vortex position $w(z)$ at the renormalized critical current $I=I_{c}$ just before depinning. The Josephson penetration depth is $\lambda_J=0.066 R$, the height of the cylinder is $h=20 R$, and in terms of the Josephson penetration depth it is $h=300\lambda_J$. The coupling strength is modulated by $m(\theta)=0.2\cos(\theta)$ with the initial position of both the flux line and the current vortex being at $\pi$. The inset shows a close-up of the region $ h-2\lambda_J < z \leq h $ where $u(z)$ and $w(z)$ differ. For most of the height, however, both lines are the same. The inset also shows the linear extrapolation of $u(z)$ after $z=h^*=h-\lambda_J$, which is used in our effective model where we take the slope at this point, $u'(h^*)$, as an effective boundary condition.
}
\end{figure}

We derive here the effective free energy functional $F^*[u(z)] $ for the displacement $u(z)$ of the spontaneous flux line. We insert the ansatz for $ \phi(\theta , z) $ from Eq.~(\ref{eq:ansatz}) into the sine-Gordon free energy from Eq.~(\ref{eq:sgfree}) including a modulation $m(\theta)$ of the coupling strength.
Averaging the different terms over $ \theta $ we find
\begin{align} 
&\left\langle \left(\frac{\partial \phi}{\partial z} \right)^2 \right\rangle_{\theta} \\ 
&= \left\langle (v')^2 + \frac{4 \Lambda u' v'}{ \cosh[ \Lambda ( \theta - u)] } + \frac{4 \Lambda^2 (u')^2}{\cosh^2[ \Lambda ( \theta - u)] } \right\rangle_{\theta} \nonumber \\ 
&\approx  (v')^2 + 2 u' v' + \frac{4 \Lambda}{\pi}  (u')^2, \nonumber
\end{align} 
and
\begin{align}
& \left\langle \left( \frac{\partial \phi}{\partial \theta} \right)^2 \right\rangle_{\theta} \\
& = \left\langle \left(1 - \frac{2 \Lambda}{ \cosh[ \Lambda ( \theta - u)] } \right)^2 \right\rangle_{\theta}\approx  \frac{4 \Lambda}{\pi} -1, \nonumber
\end{align} 
and
\begin{align}
&\left\langle \left(1+m(\theta)\right) \cos(\phi-\theta) \right\rangle_{\theta} \\
&= \left\langle (1+m(\theta)) \cos v \left(1- \frac{2}{\cosh^2[\Lambda(\theta-u)]} \right) \right\rangle_{\theta} \nonumber\\
&-\left\langle (1+m(\theta))  \sin v \frac{2 \sinh[\Lambda(\theta-u)]}{ \cosh^2[\Lambda(\theta-u)]} \right\rangle_{\theta} \nonumber\\
&\approx \cos v \left(1 - \frac{2}{\pi \Lambda} (1+ m(u)) \right) - \sin v \frac{1}{\Lambda^2} m'(u) . \nonumber
\end{align}
Collecting all terms depending on $ u(z) $ and $ v(z) $ while dropping the constants and a global prefactor we find
\begin{align}
F[u,v]    = \int_0^h\!\text{d}z\, &\left[ \frac{(u')^2}{2} + \frac{1}{2R^2}  m(u) \cos v \right.  \nonumber \\  
&      + \frac{\pi \lambda_J}{4 R^3} m'(u)  \sin v  + \frac{\pi \lambda_J}{4 R} u' v' \\ 
&   \left.   + \frac{\pi \lambda_J}{4R}  \left\{ \frac{(v')^2}{2} - \frac{1}{\lambda_J^2} \cos v \right\}  \right]  \nonumber
\label{functional-1}
\end{align}
with the boundary conditions 
\begin{align}
v'(h) = \gamma , \quad   v'(0) = 0, \\ 
u'(h)=0,  \quad u'(0) = 0 .
\end{align}
First, we consider the leading terms in the variation of $ F $ with respect to $v$,
\begin{equation}
v'' = \frac{1}{\lambda_J^2} \sin v,
\end{equation}
which is solved approximately by
\begin{equation}
v(z) = \gamma \lambda_J e^{(z-h)/\lambda_J} ,
\end{equation}
which is consistent with the guess based on the solution from the unfrustrated case in Eq.~(\ref{eq:unfrust_sol}).
This solution is confined in a very narrow region $h-\lambda_J \leq z\leq h$, where the variation of $F$ with respect to $u $ is
\begin{equation}
u'' = - \frac{\pi \lambda_J}{4 R} v''.
\end{equation}
Integrating and taking into account the boundary condition we then obtain close to $ z =h $
\begin{equation}
u'(z) = - \frac{\pi \lambda_J}{4 R} \left\{ \gamma e^{(z-h)/\lambda_J} - \gamma \right\}  .
\end{equation}
In the range $ 0 \leq z \leq h - \lambda_J \sim h^* $ we can therefore describe the behavior of $ u (z) $ using the above functional with $ v(z) \approx 0 $,
\begin{equation}
F^*[u] = \int_0^{h^*} dz \; \left[ \frac{(u')^2}{2} + \frac{1}{2 R^2} m(u) \right],
\end{equation}
and with an effective boundary condition
\begin{equation}
u'(h^*) = \frac{\pi \lambda_J}{4 R} \gamma.
\label{eq:bc_eff}
\end{equation}
The additional bending of $ u(z) $ in the range $ h^* < z \leq h $ is neglected. This approach is used in Sec.~\ref{sec:pin-strong} to discuss the pinning effect of the flux line, approximating $ h^* \approx h $. In Fig.~\ref{fig:vort} we show both the full $u(z)$ and the linear extrapolation corresponding to the effective model with $ h^* = h $.

While $u(z)$ captures the displacement of the flux line, the actual position $w(z)$ of the current vortex is given through the condition
\begin{equation}
J_r(w(z),z)=0.
\end{equation}
Our ansatz leads to
\begin{align}
w(z)&=u(z)+\lambda_J\log\left(\tan\left(\frac{v(z)+\pi}{4}\right)\right)\\
&\approx u(z)+\lambda_J\frac{v(z)}{2}.
\end{align}
The flux line displacement $u(z)$ and the current vortex $w(z)$ can be extracted from the numerical data through
\begin{equation}
u(z)=\min_{\theta}B_z(\theta,z),
\end{equation}
and indirectly through
\begin{equation}
\phi(\theta,z)=\theta+\pi\text{ at }\theta=w(z).
\end{equation}
Both $u(z)$ and $w(z)$ are plotted in Fig.~\ref{fig:vort}, highlighting the meaning of $h^*$.

Note that for solutions close to the minimum $ u= u_0 $ of $ m(u) $ there is a natural length scale $\ell$ found through the parabolic approximation
\begin{equation}
F^*[u] = \int_0^{h^*} dz \; \left[ \frac{(u')^2}{2} + \frac{1}{4 R^2} m''(u_0) u^2 \right]
\end{equation}
and given by
\begin{equation}
\ell = \sqrt{\frac{2 R^2}{ m''(u_0)}}.
\label{eq:ell}
\end{equation}
We assume that $ \lambda_J \ll \ell \ll h^*,h $. 

\bibliography{myref}

\end{document}